# High-Fidelity Single-Shot Quantitative Differential Phase Microscopy Using Pseudothermal Sagnac Interferometer

**Paweł Gocłowski[1], Hong Mao[1], Maciek Trusiak[2], Balpreet S. Ahluwalia[1,3,4,†], Azeem Ahmad[1,*, †]**

[1]*Department of Physics and Technology, UiT The Arctic University of Norway, 9037 Tromsø, Norway*
[2]*Warsaw University of Technology, Institute of Micromechanics and Photonics, 8 Sw. A. Boboli St., 02-525 Warsaw, Poland*
[3]*Department of Clinical Science, Intervention and Technology, Karolinska Institute, Stockholm, Sweden*
[4]*The Faculty of Mathematics and Natural Sciences, Department of Physics, University of Oslo, 0313 Oslo, Norway*
[†]*Shared authors.*
[*]*ahmadazeem870@gmail.com*
Email: *balpreet.singh.ahluwalia@uit.no, pawel.goclowski@gmail.com*

**Abstract: In this letter, a high-fidelity single-shot differential quantitative phase microscopy (dQPM) method is presented to effectively image nearly transparent biological samples. The proposed method is based on a common-path Sagnac interferometric configuration, which provides superior temporal phase stability and robustness against environmental disturbances. The proposed system exploits a pseudothermal source to achieve high spatial sensitivity and generate dense interference fringes for effective single-shot differential quantitative phase imaging. The effectiveness of the proposed system is experimentally demonstrated with various samples, including polystyrene microspheres, a USAF phase target, fixed and live HeLa cells, and mouse kidney tissue.**

Quantitative Phase Microscopy (QPM) is a well-established non-contact, non-invasive and label-free quantitative microscopy technique widely used for biological sample imaging. It allows high contrast observation of nearly transparent structures like cells or thin tissue sections [1]. While techniques like phase contrast and differential interference contrast (DIC) [2] are focused on pure, qualitative contrast enhancement and fluorescence microscopy is focused on providing subcellular specificity and superior contrast imaging via chemical labels, QPM utilizes interference of the object and the reference beam in order to provide precise quantitative phase information about the sample. In QPM, interferograms are recorded and post-processed to obtain the phase map of the sample. The retrieved phase information can be used to determine cell's area, volume, dry mass, etc.[3]. Tracking these parameters in a live cell via timelapse imaging is vital for observation and characterization of various biophysical processes, such as, disease tracking and drug performance investigation. Contrary to the fluorescence approaches, QPM is a label free technique and devoid of photo-bleaching issues, therefore, it is considered as more live cell friendly imaging approach.

Most of the conventional QPM solutions rely on non-common-path geometry, where object beam and reference beam follow entirely different paths to form interference fringes. Such approaches are typically based on Mach-Zehnder [4] or Linnik [5] interferometers. These configurations allow easy implementation of single-frame or multi-frame phase reconstruction and provide reliable direct phase measurements of the specimens. However, these systems are significantly prone to any kind of environmental instability, like vibrations, airflow, temperature change etc. This leads to a change in the optical path difference (OPD) between the two arms of the interferometer, which degrades the temporal phase stability [6]. To address these challenges, various common-path interferometric configurations have been developed and applied in biological imaging. In such solutions, both interfering beams follow the same optical path, minimizing environmental vulnerability and improving the temporal phase stability.

DIC stands out as one of the earliest and most known imaging techniques, which followed the principle of shearing interferometry. It relies on two slightly separated, orthogonally polarized light beams transmitted through the sample. In those beams, variation in sample's thickness or refractive index is encoded as a phase difference. As a result, high contrast image is obtained with high spatial resolution, enhanced edges and the impression of three-dimensional object. DIC method has been widely applied to biological studies thanks to its simplicity and compatibility with brightfield microscopes. However, the original DIC approach is inherently qualitative.

Over the years, DIC concept has been modified and turned into fully quantitative via approaches like retardation-modulated DIC [7], phase-shifted DIC (PS-DIC) [8], gradient light interference microscopy (GLIM) [9], quadriwave lateral shearing interferometry (QWLSI) [10], calcite beam displacer based quantitative DIC [11, 12], and others. Common property of these approaches is that through small shear between the beams (typically half of the diffraction limit), they provide information about the first derivative of the phase instead of direct phase measurement. Differential measurement allows for easier highlight of the subcellular details and easier biological interpretation of the result. Moreover, shearing interferometry does not require artifact-prone stage of phase unwrapping as the first derivative of the phase typically does not exceed 2π[13-15], at least as long as the sample is continuous and not step-like.

In this work, a new single-shot differential quantitative phase microscopy (dQPM) approach is presented, employing a Sagnac interferometer–based common-path configuration and a pseudothermal light source to achieve high temporal phase stability and enhanced spatial phase sensitivity. Pseudothermal light source allows acquisition of speckle free and low noise differential phase images comparable to low coherence interferometry [5, 16]. Sagnac configuration allows for convenient adjustment of the shear distance, density of the interferometric fringes and maintaining exquisite temporal stability of the registered data. The control of the interference fringe density makes the proposed configuration compatible with different cameras of different pixel sizes. In addition, through the polarizer rotation, interference fringes can be washed out to generate reference bright field image. The proposed dQPM system operates in single shot mode and provides diffraction limited resolution in the reconstructed differential phase maps, making it suitable for the investigation of dynamic biological phenomena. The potential of the proposed system is demonstrated on bead samples, USAF phase chart with varying heights, both fixed and living HeLa cells and mouse kidney tissues.

For the dQPM configuration, the two interfering fields are identical replicas of the object's wavefront, which is symmetrically displaced with respect to the optical axis. If the total shear between the two wavefronts is represented by d, then the two fields are sheared by $\pm d/2$. The two fields also have opposing tilt angles of $-\theta_x$ and $+\theta_x$, which give rise to a spatial carrier along the x-direction. The complex fields of the two interfering beams can thus be represented by

$$E_1 = A_1 \exp\left[i\left(\phi\left(x - \frac{d}{2}, y\right) - kx \sin \theta_x\right)\right] \tag{1}$$

$$E_2 = A_2 \exp\left[i\left(\phi\left(x + \frac{d}{2}, y\right) + kx \sin \theta_x\right)\right] \tag{2}$$

where $k = 2\pi/\lambda$ and $\phi(x, y)$ denotes the object phase. The recorded interferogram is given by $I = |E_1 + E_2|^2$, which results in

$$I = |A_1|^2 + |A_2|^2 + 2A_1 A_2 \cos\left[2kx \sin\theta_x + d\frac{\partial\phi}{\partial x}\right] \tag{3}$$

Defining

$$A_0 = |A_1|^2 + |A_2|^2, \tag{4}$$

and the fringe visibility

$$b = \frac{2A_1 A_2}{|A_1|^2 + |A_2|^2}, \tag{5}$$

the interferogram can be expressed in normalized form as

$$I = A_0\left[1 + b \cos\left(2kx\sin\theta_x + d\frac{\partial\phi}{\partial x}\right)\right] \tag{6}$$

Since, the carrier frequency is defined as

$$f_x = \frac{2\sin\theta_x}{\lambda} \tag{7}$$

Therefore, the recorded intensity can be written as

$$I = A_0\left[1 + b \cos\left(2\pi f_x x + d\frac{\partial\phi}{\partial x}\right)\right] \tag{8}$$

This expression shows that the interferogram encodes the first spatial derivative of the object phase scaled by the total shear $d$. Detailed mathematical derivations are provided in the Supplementary Material.

To retrieve the differential phase of the object, single-shot Fourier Transform method is utilized [17]. The steps for differential phase recovery are shown in Fig. 1(a)–1(c). The interferogram, its Fourier spectrum, and the recovered differential phase map obtained by filtering the +1 diffraction order and applying an inverse Fourier transform are illustrated in Fig. 1(a)–1(c), respectively. Note that high-frequency carrier fringes are generated to separate information peaks in the Fourier domain which is required for diffraction limited phase recovery. For each sample, one reference interferogram is acquired in a sample-free region. The differential phase maps recovered from the object and the reference interferograms are then subtracted from each other to remove unwanted background variations.

Figure 1(d) shows the schematic diagram of the proposed dQPM approach. The laser beam (638 nm diode laser, OptLasers) is first passed through a rotating diffuser (RD) to reduce spatial coherence while preserving temporal coherence. Such light source provides speckle-free illumination and high-quality interference fringes at the same time as long as the speckle fields in both interfering beams are mutually correlated [18]. The output of RD is collimated by lens $L_1$ and polarized by a linear polarizer $P_1$ oriented at a 45-degree angle.

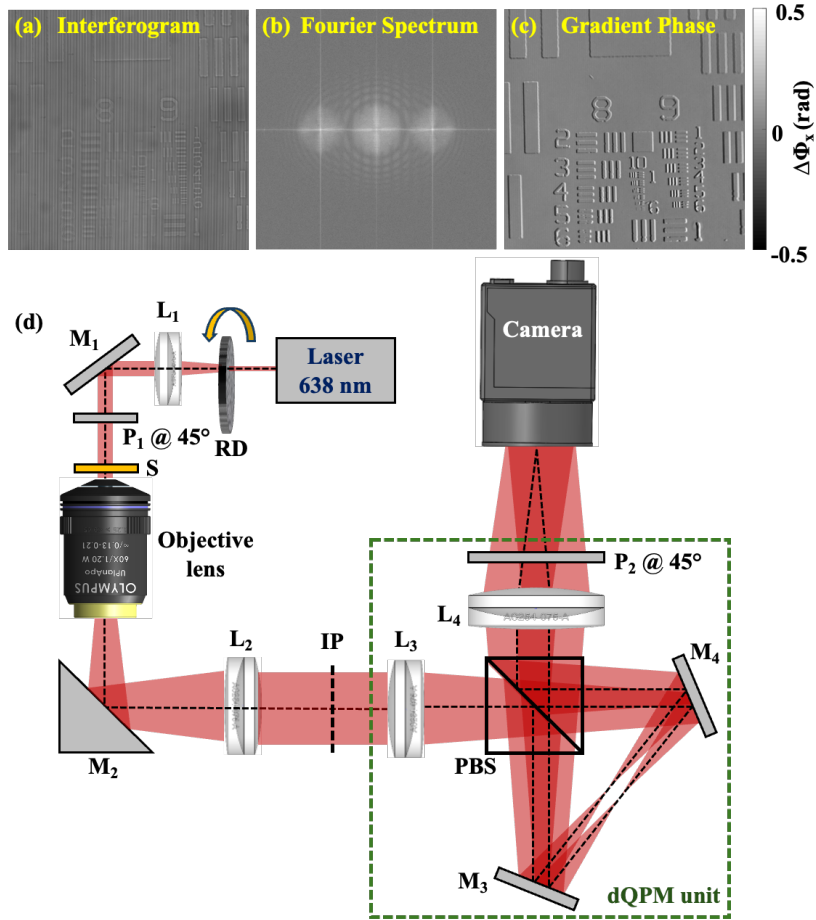

*Figure 1 - (a) Interferogram, (b) Fourier spectrum, and (c) recovered differential phase map of phase USAF chart. (d) Schematic of the proposed differential quantitative phase microscopy (dQPM) system.*

The setup works in the inverted transmission mode, which ensures compatibility with standard biological sample preparation substrates such as glass slides and petri dishes. Objective lens collects information about the sample and forms image at the image plane (IP) via lens $L_2$. The image is then relayed to the camera plane via a 4-f system consisting of lens $L_3$ and $L_4$. Two replicas of the object beam are generated by Sagnac interferometric configuration, placed between $L_3$ and $L_4$ (see Fig.1d).

Sagnac interferometer consists of one polarizing beam splitter (PBS) and two mirrors ($M_3$ and $M_4$). The PBS splits the input beam into two orthogonally polarized beams. The transmitted and reflected beams are directed towards the same PBS via mirrors $M_3$ and $M_4$ forming a cyclic path.

The transmitted and reflected polarization components get transmitted and reflected again through the same PBS with minimal loss of intensity. As a result, we get two parallel propagating output beams with orthogonal polarization states. Tilt of $M_3$ and $M_4$ controls the separation between the output beams, thereby controlling the angle between the beams when combined via lens $L_4$. $M_3$ and $M_4$ enable appropriate adjustment of fringe density and orientation. At the end, polarization state of the beams is unified by polarizer $P_2$ oriented at 45° to obtain high contrast interference fringes at the detector. The polarizer's pass axis can be oriented either vertically or horizontally to wash out the interference fringes, thus forming a brightfield image.

Note that due to the common path geometry of dQPM, any objective lenses can be freely used without disturbing the optical configuration. However, in this work all the results were acquired with 60×/1.2NA water immersion objective lens from Olympus. All interferograms were acquired with Hamamatsu Orca Flash 4.0 CMOS camera. The camera has 2048 × 2048 pixels with pixel size of 6.5 µm. The total magnification of the system is measured to be equal to 122× by imaging standard amplitude USAF test, resulting in 53 nm pixel size at the sample plane. The field of view (FoV) of the setup is equal to 109 × 109 µm.

First, we quantified the spatial and temporal phase sensitivity of the system, which were measured to be 6 mrad and 12 mrad, respectively (see Supplementary Fig. S1 for details). Second, the optimal shear required for diffraction-limited and low-noise differential phase recovery was systematically investigated. The camera was mounted on a 1D translation stage to control the shear between the overlapped wavefronts. After each camera translation, the sample image was refocused using the sample's piezo stage to avoid defocus. It was observed that small shear values introduce significant noise, whereas large shear values reduce the spatial resolution of the recovered differential phase maps. Therefore, a series of interferograms were acquired at different shear values by translating the camera to determine the optimal shear for subsequent experiments.

First, the zero-shear position was approximately determined by optimally overlapping the two replicas of a sample image (e.g., beads). Subsequently, the interferogram was recorded and the corresponding differential phase maps were reconstructed. At the zero-shear position, the phase of the overlapped wavefront replicas nearly cancels the specimen phase. The camera was then translated from - 4.5 mm to + 4.5 mm in 0.2 mm increments to record the corresponding interferograms. This was followed by phase reconstruction to determine the optimal camera position (i.e., shear value). It was found that a camera displacement of 2 mm from the zero-shear position produced a diffraction limited differential phase image with a high signal-to-background ratio.

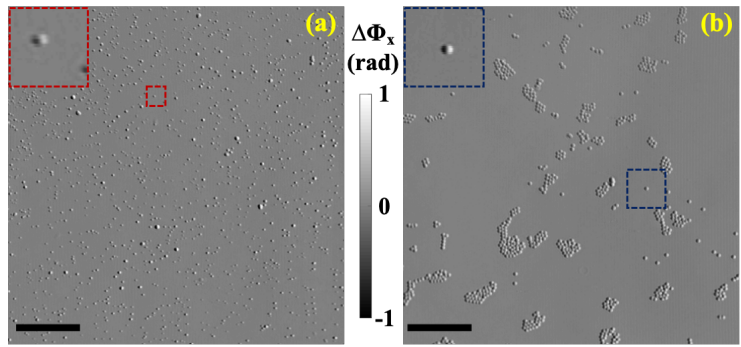

*Figure 2 - Differential phase reconstruction of nanobeads: 300 nm (a) and 1 µm (b) in diameter. Zoomed views of a single bead are shown in dashed red and blue rectangles. The scale bars are equal to 20 µm.*

Figures 2(a) and 2(b) show the reconstructed differential phase of 300 nm and 1 µm bead sample, respectively. The recovered phase maps of a single bead at different camera positions, along with the corresponding line profiles, are shown in Supplementary Fig. S2. We further investigated shear variations at the 2 mm camera position across four different locations within the field of view (FOV). A slight variation in shear was observed, likely due to distortion introduced by the tube lens ($L_4$) positioned before the camera.

This issue can be mitigated in future implementations by using an ultra-low-distortion tube lens. The shear analysis of four different FOV positions is provided in the Supplementary Information.

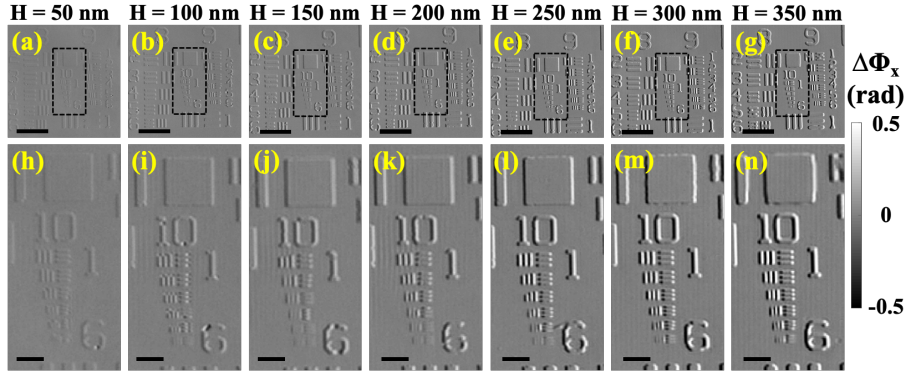

*Figure 3 - Differential phase imaging of phase USAF test. Resolution tests of 50 (a), 100 (b), 150 (c), 200 (d), 250 (e), 300 (f) and 350 nm depth (g). Corresponding zoomed views of the 10th group (h-n) highlighted with black dashed rectangles (a-g). The scale bars are equal to 20 µm (a-g) and 5 µm (h-n).*

Subsequently, the optimized dQPM setup was implemented to evaluate its spatial resolution performance and phase sensitivity by conducting experiments on a phase USAF test target (Benchmark Technologies) with seven different heights. Figures 3(a) – 3(g) present the differential phase maps of the USAF target with depths ranging from 50 to 350 nm in 50 nm increments. The magnified views of the regions marked with black dotted boxes are shown in Figs. 3(h)–3(n). The results demonstrate that the phase sensitivity of our system is better than 50 nm (Fig. 3(h)). The system resolved up to 6th element of the 10th group (which corresponds to 548 nm full pitch lateral resolution), which is found to be in good agreement with the theoretical value of 490 nm. The reconstructions of the 5th and 6th elements for structures shallower than 200 nm exhibit a loss of spatial resolution (Figs. 3(h)–3(j)). This may be due to fabrication-related imperfections, as closer inspection of the reconstructed differential phase maps reveals such effects in the thinner features of the resolution target. However, systematic profilometric measurements are required to confirm this, and these will be conducted in future studies.

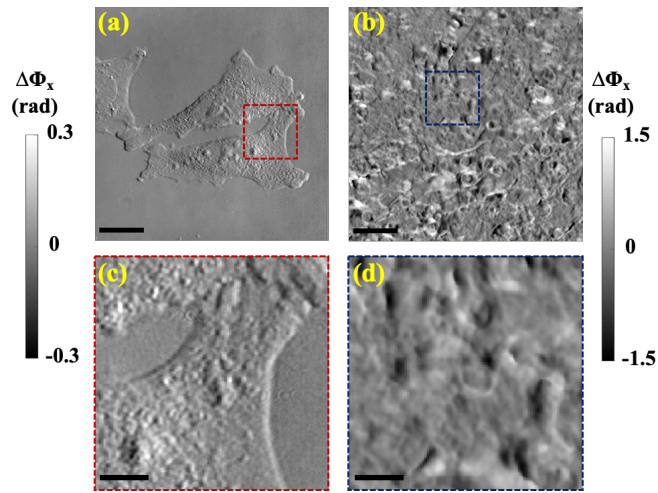

*Figure 4 - Differential phase imaging of fixed HeLa cells (a) and mouse kidney tissue (b). Zoomed views of smaller regions (c, d) highlighted with red and blue dashed rectangles, respectively. The scale bars are equal to 20 µm (a, b) and 5 µm (c, d).*

Next, performance of the proposed dQPM system has been tested on biological specimens. Figure 4 shows the differential phase reconstructions of two fixed samples: HeLa cells (Fig. 4(a)) and mouse kidney tissue (Fig. 4(b)). Overall performance does not differ from the test objects presented in Fig. 2 and 3. Note that the proposed approach helps to achieve high structural contrast and enhance visibility of subcellular details in a similar way to DIC, but with quantitative information. Enhancement of cell edges may enable easier tracking of cell shape and area. Figs. 4(c) and 4(d) show zoomed regions of the specimens.

To further demonstrate the practical potential of dQPM, experiments were performed on live HeLa cells. A 70-minute time-lapse sequence was recorded with time interval of 1 minute per frame (Visualization 1). Figure 5 presents a representative interferogram and the corresponding differential phase map extracted from the time-lapse movie. A closer examination of the time-lapse sequence reveals dynamic morphological changes in HeLa cells after removal from the incubator and placement under the microscope, where temperature and humidity conditions were not fully controlled. Under these suboptimal environmental conditions, the cells exhibit stress-related responses and apparent intercellular interactions. A more detailed analysis of the live-cell experiment, along with the complete time-lapse frames and video, is provided in the Supplementary Information (Visualization 2).

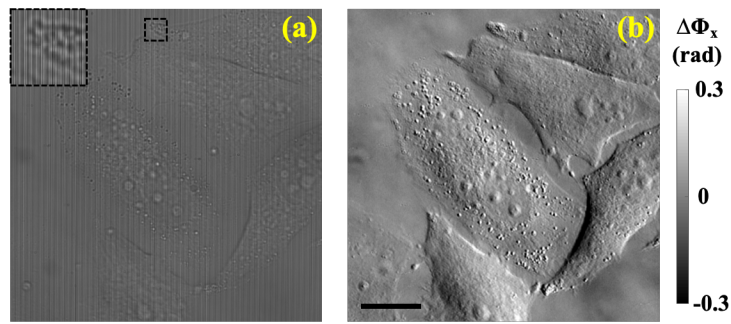

*Figure 5 - Live HeLa cells imaging. Exemplary interferogram (a) and differential phase map (b). The scale bar is equal to 20 µm.*

In this study, we have proposed a new approach to differential QPM. The system is found to be well suited for broad variety of standard and biological sample imaging. It's common-path architecture based on Sagnac interferometer ensures high temporal stability of 12.2 mrad. and robustness to environmental disturbances. In addition, reduced spatial coherence of the light source enhances spatial phase sensitivity and is found to be equal to 6 mrad., which allows for studying the thin biological specimens. Moreover, its single-shot phase reconstruction capability provides high temporal resolution suitable for dynamic sample observation.

The proposed system employs a shearing interferometric configuration in which two spatially separated wavefronts propagate through the tube lens at off-axis positions. While this approach enables differential phase measurement with high sensitivity, it also makes the system particularly susceptible to optical distortions introduced by the tube lens. Since the interfering wavefronts do not travel along the optical axis, any asymmetric distortion imposed by the tube lens affects the wavefronts differently, leading to systematic errors that may be misinterpreted as sample-induced phase variations. Therefore, the optical quality and alignment precision of the tube lens become critical factors in maintaining measurement accuracy. It is observed that the distortion of the tube lens causes non-constant shear throughout the FoV, and causes different differential phase values in the middle and the corner regions of the FoV. This issue can be fixed by incorporating ultralow distortion tube lens in the proposed scheme.

In addition, the proposed scheme can be extended into a full QPM technique by integrating two Sagnac interferometers to measure differential phase maps along orthogonal X and Y directions. This would facilitate robust phase integration. Furthermore, the system could be extended toward three-dimensional refractive index tomography with the relatively simple addition of rotating oblique illumination, enabling volumetric reconstruction without substantial modifications to the existing setup.

**Data availability statements.** The datasets generated and/or analyzed during the current study are available from the corresponding author upon reasonable request.

**Funding.** A.A. acknowledges FRIPRO Young (project # 345136) funding from Research Council of Norway. B.S.A. and P.G. acknowledges the funding from UiT Thematic Funding (NASAR). B.S.A. acknowledges funding received from the European Union's HORIZON Research and Innovation Actions under grant agreement No. 101191315. M.T. acknowledges project # 2023/48/Q/ST7/00172 funding from Narodowe Centrum Nauki. The publication charges for this article have been funded by a grant from the publication fund of UiT The Arctic University of Norway.

**Author Contribution.** A.A. conceptualized the idea and designed the optical configuration. A.A. and B.S.A supervised and conceived the project. P.G and A.A. developed the experimental setup. P.G. performed the experiments, analyzed the data, and prepared the figures. P.G. and A.A. mainly wrote the manuscript. H.M. prepared the biological samples and provided biological insights. M.T. contributed to the insights into dQPM. All authors reviewed and edited the manuscript.

**Conflict of interest.** Azeem Ahmad and Balpreet Singh Ahluwalia have submitted a patent application to protect the invention of dQPM (Patent Application No. 2513325.7). All other authors declare no competing interests

**Acknowledgements.** We would like acknowledge Prof. Kristin Andreassen Fenton for providing mouse kidney tissue sample.

**Supplemental document.** See Supplement 1 for supporting content.

6. Shemonski, N.D., et al., *Stability in computed optical interferometric tomography (Part I): Stability requirements.* Opt. Express, 2014. **22**(16): p. 19183-19197.
7. Ishiwata, H., M. Itoh, and T. Yatagai, *A new analysis for extending the measurement range of the retardation-modulated differential interference contrast (RM-DIC) microscope.* Opt. Commun., 2008. **281**(6): p. 1412-1423.
8. King, S.V., et al., *Quantitative phase microscopy through differential interference imaging.* J. Biomed. Opt., 2008. **13**(2): p. 024020-024020-10.
9. Nguyen, T.H., et al., *Gradient light interference microscopy for 3D imaging of unlabeled specimens.* Nat. Commun. , 2017. **8**(1): p. 210.
10. Yuan, X., et al., *Optical diffraction tomography based on quadriwave lateral shearing interferometry.* Opt. Laser Technol., 2024. **177**: p. 111124.
11. Kou, S.S. and C.J. Sheppard. *Quantitative phase restoration in differential interference contrast (DIC) microscopy*. in *Opt. Digit. Image Process.* 2008. SPIE.
12. Saxena, A., et al., *Single-shot quantitative differential phase contrast microscope using a single calcite beam displacer.* Appl. Opt., 2024. **63**(32): p. 8350-8358.
13. Yuan, X., et al., *High-precision gaseous flame temperature field measurement based on quadriwave-lateral shearing interferometry.* Optics and Lasers in Engineering, 2023. **162**: p. 107430.
14. Yuan, X., et al., *Optical diffraction tomography based on quadriwave lateral shearing interferometry.* Optics & Laser Technology, 2024. **177**: p. 111124.
15. Baffou, G., *Wavefront microscopy using quadriwave lateral shearing interferometry: from bioimaging to nanophotonics.* ACS photonics, 2023. **10**(2): p. 322-339.
16. Ahmad, A., et al., *High-throughput spatial sensitive quantitative phase microscopy using low spatial and high temporal coherent illumination.* Scientific reports, 2021. **11**(1): p. 15850.
17. Takeda, M., H. Ina, and S. Kobayashi, *Fourier-transform method of fringe-pattern analysis for computer-based topography and interferometry.* J. Opt. Soc. Am. A, 1982. **72**(1): p. 156-160.
18. Ahmad, A., N. Jayakumar, and B.S. Ahluwalia, *Demystifying speckle field interference microscopy.* Sci. Rep., 2022. **12**(1): p. 10869.
9

# High-Fidelity Single-Shot Quantitative Differential Phase Microscopy Using Pseudothermal Sagnac Interferometer

**Paweł Gocłowski[1], Hong Mao[1], Maciek Trusiak[2], Balpreet S. Ahluwalia[1,3,4,†], Azeem Ahmad[1,\*, †]**

[1]Department of Physics and Technology, UiT The Arctic University of Norway, 9037 Tromsø, Norway
[2]Warsaw University of Technology, Institute of Micromechanics and Photonics, 8 Sw. A. Boboli St., 02-525 Warsaw, Poland
[3]Department of Clinical Science, Intervention and Technology, Karolinska Institute, Stockholm, Sweden
[4]The Faculty of Mathematics and Natural Sciences, Department of Physics, University of Oslo, 0313 Oslo, Norway
[†]Shared authors.
[*]ahmadazeem870@gmail.com
Email: balpreet.singh.ahluwalia@uit.no, pawel.goclowski@gmail.com

**Visualization 1:** Time-lapse interferometric recording of live HeLa cells.

**Visualization 2:** Time-lapse quantitative differential phase map of live HeLa cells.

**Visualization 3:** Time-lapse sample free interferometric recording for spatial and temporal phase stability.

**Mathematical Description:** In dQPM, two identical wavefronts are superimposed after symmetrically displacing them relative to the optical axis. In addition, if the total shear is d, each field is shifted by $\pm d/2$. Furthermore, the two fields are propagating with opposite tilt angles $-\theta_x$ and $+\theta_x$. Under these circumstances, the complex optical fields can be expressed as follows.

$$E_1 = A_1 \exp\left[i\left(\phi\left(x - \frac{d}{2}, y\right) - kx\sin\theta_x\right)\right] \quad (1)$$

$$E_2 = A_2 \exp\left[i\left(\phi\left(x + \frac{d}{2}, y\right) + kx\sin\theta_x\right)\right] \quad (2)$$

where

$$k = 2\pi/\lambda$$

The interference intensity is

$$I = |E_1 + E_2|^2 \quad (3)$$

which expands to

$$I = |E_1|^2 + |E_2|^2 + E_1^* E_2 + E_1 E_2^* \quad (4)$$

Now,

$$|E_1|^2 = |A_1|^2, |E_2|^2 = |A_2|^2$$

Next, compute the cross term:

$$E_1^* = A_1 \exp\left[-i\left(\phi\left(x - \frac{d}{2}, y\right) - kx\sin\theta_x\right)\right]$$

So

$$E_1^* E_2 = A_1 A_2 \exp\left\{i\left[\phi\left(x + \frac{d}{2}, y\right) - \phi\left(x - \frac{d}{2}, y\right) + 2kx\sin\theta_x\right]\right\} \quad (5)$$

Similarly,

$$E_1 E_2^* = A_1 A_2 \exp\left\{-i\left[\phi\left(x+\frac{d}{2},y\right) - \phi\left(x-\frac{d}{2},y\right) + 2kx \sin \theta_x\right]\right\} \quad (6)$$

Using

$$e^{i\alpha} + e^{-i\alpha} = 2\cos \alpha,$$

the sum of the cross terms becomes

$$E_1^* E_2 + E_1 E_2^* = 2A_1 A_2 \cos\left[\phi\left(x+\frac{d}{2},y\right) - \phi\left(x-\frac{d}{2},y\right) + 2kx \sin \theta_x\right] \quad (7)$$

Therefore,

$$I = |A_1|^2 + |A_2|^2 + 2A_1 A_2 \cos\left[\phi\left(x+\frac{d}{2},y\right) - \phi\left(x-\frac{d}{2},y\right) + 2kx \sin \theta_x\right] \quad (8)$$

Now apply the first-order Taylor expansion about $x$:

$$\phi\left(x+\frac{d}{2},y\right) \approx \phi(x,y) + \frac{d}{2}\frac{\partial \phi}{\partial x} \quad (9)$$

$$\phi\left(x-\frac{d}{2},y\right) \approx \phi(x,y) - \frac{d}{2}\frac{\partial \phi}{\partial x} \quad (10)$$

Subtracting Eq. (10) from Eq. (9),

$$\phi\left(x+\frac{d}{2},y\right) - \phi\left(x-\frac{d}{2},y\right) \approx d \frac{\partial \phi}{\partial x} \quad (11)$$

Substituting this into Eq. (8) gives

$$I = |A_1|^2 + |A_2|^2 + 2A_1 A_2 \cos\left[2kx \sin \theta_x + d \frac{\partial \phi}{\partial x}\right] \quad (12)$$

Define

$$A_0 = |A_1|^2 + |A_2|^2 \quad (13)$$

and

$$b = \frac{2A_1 A_2}{|A_1|^2 + |A_2|^2} \quad (14)$$

Then Eq. (12) becomes

$$I = A_0 \left[1 + b \cos\left(2kx\sin \theta_x + d \frac{\partial \phi}{\partial x}\right)\right] \quad (15)$$

Since

$$2k\sin \theta_x = 2\left(\frac{2\pi}{\lambda}\right)\sin \theta_x = 2\pi f_x,$$

the carrier frequency is

$$f_x = \frac{2\sin \theta_x}{\lambda} \quad (16)$$

Hence the intensity can be written as

$$I = A_0 \left[1 + b \cos\left(2\pi f_x x + d \frac{\partial \phi}{\partial x}\right)\right] \quad (17)$$

**Phase stability:** It is expected from common-path interferometers to achieve better temporal phase stability than their non-common path counterparts. Phase stability of the proposed dQPM setup has been checked by recording 60-second-long video without any sample (Visualization 3), which in theory should correspond to phase values equal to zero. The second objective of the measurement was to exploit benefits of single-shot phase reconstruction by checking the system's capability of working with highly dynamic specimens, therefore the video has been recorded in 60 frames per second giving 3600 frames in total. The results of this study are presented in Fig. S1.

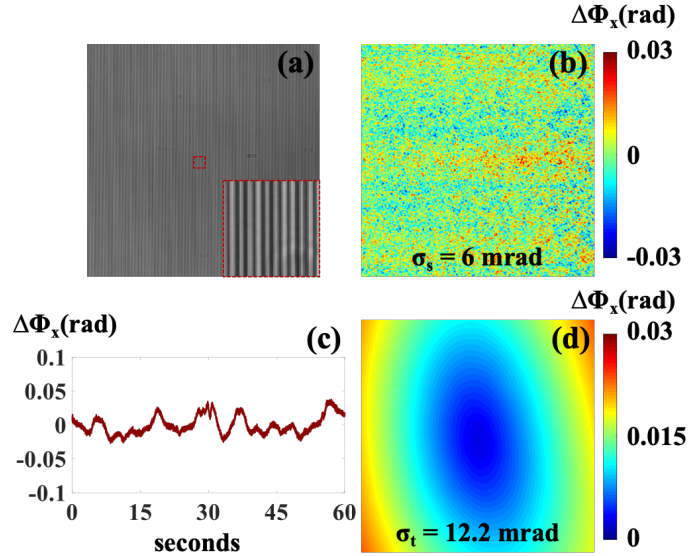

*Figure S1 - Phase stability investigation. Interferogram (a) and differential phase (b) of exemplary frame from sample-free movie. Time profile of differential phase for exemplary pixel (c). Phase stability map (d).*

One of the interferograms is shown in Fig. S1(a), and the corresponding retrieved phase map is presented in Fig. S1(b). Spatial phase stability is equal to 6 mrad and has been calculated as standard deviation of exemplary phase map from Fig. S1(b). To evaluate temporal phase stability, the phase was reconstructed from each interferogram in the time-lapse sequence. The reconstructed phase maps were stacked to form a three-dimensional matrix, where the x and y dimensions represent spatial coordinates and the z dimension corresponds to time. For each spatial location, the temporal phase profile was extracted and averaged to obtain a global temporal phase trend. This average profile was then subtracted from the temporal phase profile of each pixel to remove global phase errors arising from sample drift or microscope instability. An example of the resulting corrected temporal phase profile is shown in Fig. S1(c). Standard deviation of the line profile corresponding to every pixel is calculated and plotted together as temporal phase stability map (Figure S1c).

Typical temporal phase stability of a non-common path interferometer oscillates around 30 mrad. In dQPM phase stability is measured as 12.2 mrad which is more than double improvement. However, it is important to note that phase stability significantly varies within the FoV. Edges of the image have phase stability of 15-25 mrad, while in the center phase stability is less than 5 mrad (dark blue area in Fig. S1(d)).

It is possible to express both phase stability parameters in nanometers through comparison with objects of known height, such as 50 nm high phase USAF test presented in Figure 3h. When a line profile is taken through vertical bars of the 1st element of the 6th group, peak-to-background phase value related to 50 nm high structure has been measured as 86.5 mrad. It allows for a rough estimate that temporal phase stability of 12.2 mrad is equal to 4.76 nm and spatial phase sensitivity of 6 mrad is equal to 2.34 nm.

**Shear evaluation:** Next, we performed systematic shear investigation on standard polystyrene bead samples. Panels a1 to a16 in Figure S2 show differential phase reconstruction of a single 300 nm bead for 16 different camera positions: from – 4.5 to + 4.5 mm, with 0.6 mm increment, where 0 represents arbitrary determined zero-shear position. Panels b1 to b16 show the same analysis for 1 µm beads respectively.

It was determined based on signal to background ratio that camera positions closer than 1.5 mm to zero-shear position should be avoided (panels a7-a10 and b7-b10). The results also show that the shear value is not equal throughout the entire FoV, most likely due to slight distortion of the setup. Therefore, we have chosen to place the camera 2 mm from the zero-shear position to achieve sufficient shear over the entire FoV. More quantitative approach, which includes shear calculation and line profiles through the beads, is presented in Fig. S3. Figure S3 shows series of line profiles through 300 nm and 1 µm beads. For each sample, 16 sets of line profiles are plotted for the exact same 16 camera positions as in Fig. S2.

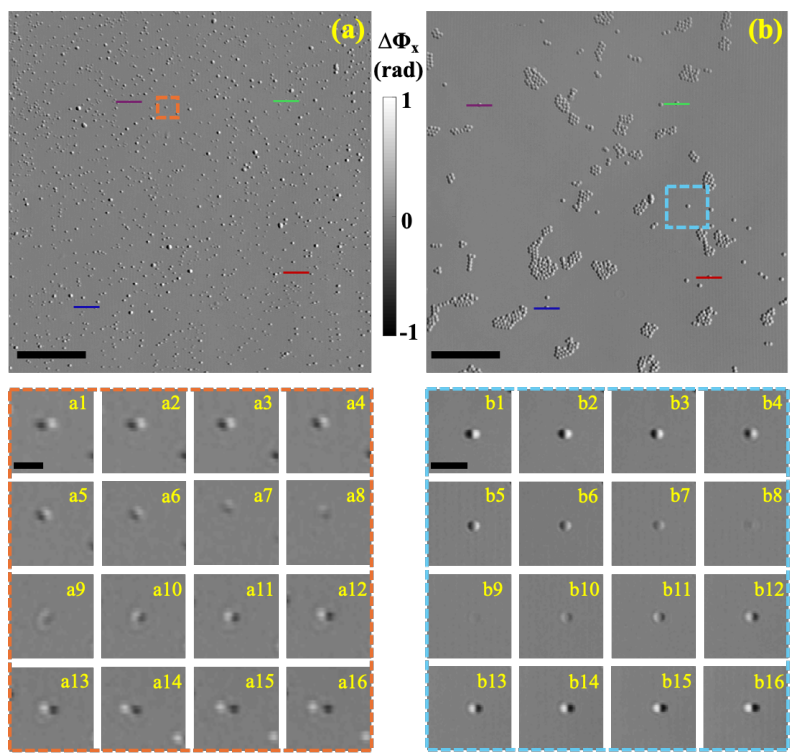

*Figure S2 - Differential phase reconstruction of nanobeads: 300 nm (a) and 1 µm (b) in diameter. Panels a1-a16 and b1-b16 show zoomed views of singular beads highlighted with orange and cyan dashed rectangles, respectively. The panels represent broad range of the camera positions – from -4,5 mm (a1, b1) to +4,5 mm (a16, b16) with 0,6 mm increment. The scale bars are equal to 2 µm (a1), 5 µm (b1) and 20 µm (a, b).*

Each set contains information about 4 beads from different corners of the image. These beads are highlighted with purple, green, blue and red solid lines in Figure S2. This is to show how the shear varies throughout the entire FoV. For enhanced readability, line profiles are separated by 1-radian offset. For each camera position, the differential phase profiles of four beads corresponding to 300 nm and 1 µm beads are plotted in Fig. S3(a) and Fig. S3(b), respectively.

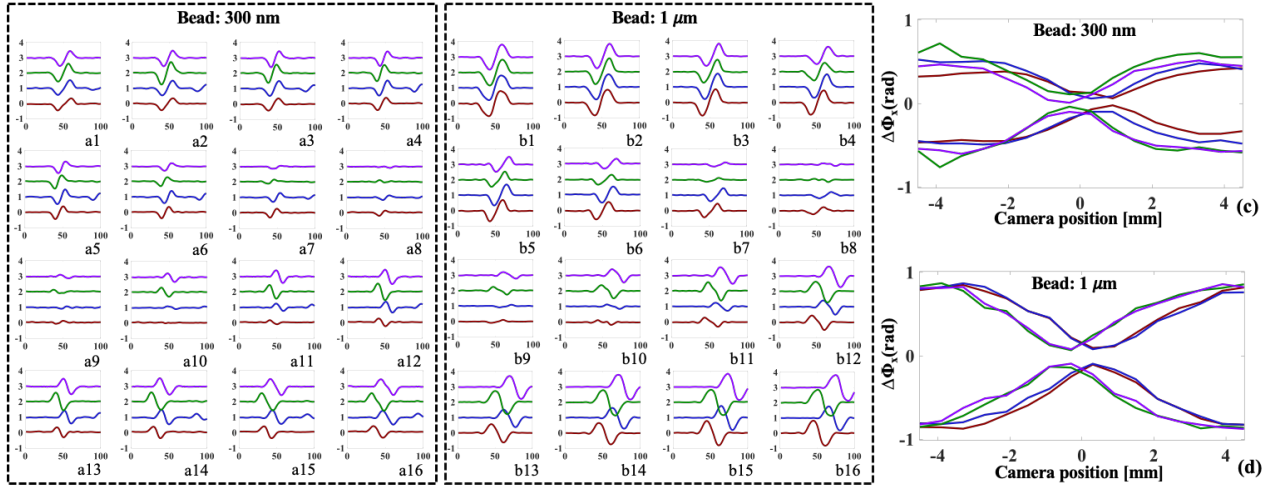

*Figure S3 - Quantitative investigation of the shear. Panels a1-a16 and b1-b16 show line profiles through 4 selected beads of 300 nm and 1 µm marked with red, blue, green and purple solid lines in Fig. S2. Colors of the line profiles match colors of the lines. The panels represent broad range of the camera positions – from -4.5 mm (a1, b1) to +4.5 mm (a16, b16) with 0.6 mm increment. The units of each line profile are differential phase in radians for Y-axis and number of pixels for X-axis. For enhanced readability, line profiles are separated by 1-radian offset. Total length of all line profiles is 100 pixels, which corresponds to 5.3 µm. For every panel, the lowest and highest peak of the line profile were recorded and plotted together (a, b). (c, d) Peak to valley difference of the differential phase profiles for 300 nm and 1 µm beads as a function of the camera positions, respectively.*

The plot confirms that the zero-shear position has been correctly determined. For each bead, the zero-shear position corresponds to the camera position where the difference between the maximum and minimum values of the line profile is minimal. It is also observed that these positions vary across the field of view. The plot shows that the 1 µm beads marked in purple and green exhibit zero shear at camera positions below zero, whereas the beads marked in red and blue exhibit zero shear at positions above zero as shown in Fig. S3(d). Therefore, the selected general zero-shear position represents an optimal compromise.

The optimal shear for subsequent experiments was determined from the line profile behavior. The shear should be as small as possible while ensuring that all four line profiles from different regions of the field of view exhibit clearly distinguishable upper and lower peaks. Based on this criterion, the optimal camera position was determined to be 2 mm from the zero-shear position, corresponding to panels a5, a12, b5, and b12 in Fig. S3.

After selecting the camera position, we attempted to validate it ultimately by measuring the shear. Theoretically, the exact determination of the shear value can be done by measuring the distance between the upper and the lower peak of the line profile. However, this approach is unreliable because the peak-to-peak separation remains nearly constant with changes in camera position. To overcome this challenge, we utilized extreme camera position from panel a15 in Fig. S3. There, 300 nm beads are fully separated and enter a total-shear regime, and peak-to-peak separation measurement is reliable for shear determination. Calculated shear is 17 pixels large which corresponds to 901 nm. Panel a15 corresponds to camera position 4.5 mm away from the zero-shear position. Therefore, shear distance d for 2 mm camera position can be expressed as:

$$d = \frac{2}{4.5} \times 901 \, nm = 400 \, nm$$

It is important to note that this calculation has been performed for exemplary bead in the bottom right part of the image, where shear is the largest and it is expected that in other parts of the FoV shear will be even smaller. In comparison, theoretical diffraction limit of our optical system is equal to:

$$diff\ limit = \frac{\lambda}{NA_{obj} + NA_{ill}} = \frac{638\ nm}{1.2 + 0.1} = 490\ nm$$

Where $NA_{obj}$ and $Na_{ill}$ denote numerical aperture of objective and illumination, respectively, while $\lambda$ represents the wavelength. The experimental resolution limit has been measured as 548 nm during investigation of phase USAF test. All this combined information proves that the shear is smaller than both theoretical and experimental resolutions of the optical system and hence should not cause resolution loss.

**Live cell investigation:** To further investigate the system performance and confirm its temporal stability, we performed a timelapse observation of live HeLa cells. Images were taken every 1 minute for 70 minutes in total. Supplementary video S2 shows differential phase maps for the entire timelapse of the whole FoV. We have carefully studied the video in search of interesting biological phenomena. Apart from dynamic movement of intracellular vesicles inside a single cell, we have also spotted interesting intercellular interaction. Left cell is transferring its resources to the other cell. Figure S4 highlights this interaction and tracks it for 35 minutes. For the purpose of the measurement, the cells were pulled out of the incubator and placed under the microscope in non-optimal conditions. When exposed to room temperature and lower humidity, cells start to struggle. Struggling cells tend to help each other survive by exchanging resources, just like in the presented observation.

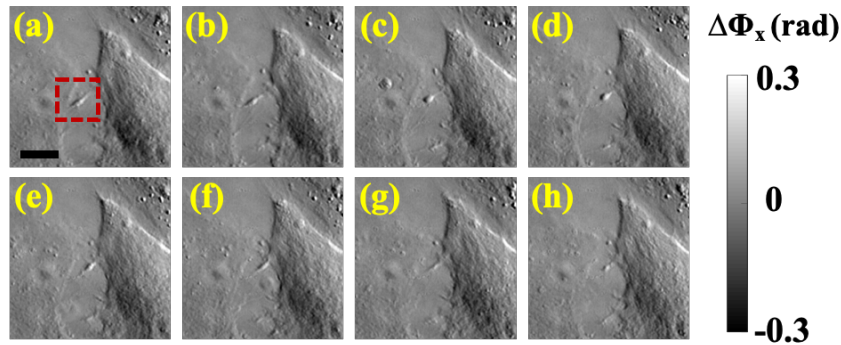

*Figure S4 - Live HeLa cell timelapse. Panels a-h show differential phase results for 8 different timepoints separated by 5 min increments. The scale bar is equal to 10 μm. Red dashed rectangle highlights resource transfer between two cells.*